\documentclass{pss}
\usepackage{graphicx}
\usepackage{amsmath}

\def\SUBMITname{phys. stat. sol. (b) \textbf{230}, No. 2, 419--423 (2002)}
\renewcommand{\figurename}{Fig.}

\begin{document}

\title{Coherence and Decoherence in Tunneling between Quantum Dots}
\author{D. M. Cardamone\thanks{) Corresponding author; e-mail: 
dmcard@physics.arizona.edu}~), C. A. Stafford, \and B. R. Barrett}
\address{Physics Department, University of Arizona, 1118 East 4th Street, 
Tucson, Arizona 85721, USA}

\submitted{}

\maketitle


\begin{abstract}
Coupled quantum dots are an example of the ubiquitous quantum double potential 
well. In a typical transport experiment,
each quantum dot is also coupled to
a continuum of states. Our approach takes this into account by using a Green's 
function formalism to solve the full system. The time-dependent solution is then
explored in different limiting cases. In general, a combination of coherent and 
incoherent behavior is observed. In the case that the coupling of each dot to 
the macroscopic world is equal, however, the time evolution is purely coherent.
\end{abstract}

The double-well potential is one of the simplest and best understood problems in
modern quantum mechanics. Its utility is likewise unparalleled. 
Potential applications of double-well devices have been noted in Refs.\
\nocite{sci95, prl98, nat98, sci98, sci01}[1-5]. For such devices to be useful, 
an understanding of the processes which couple the microscopic device 
to the macroscopic environment is paramount. That is to say, the decoherence 
processes of such systems must be well understood. To this end, we consider a 
simple exactly solvable 
model of two coupled quantum dots. An excellent review of related systems 
and some approximate solutions are given in Ref.\ \cite{leg}.

\begin{figure}[t]
\includegraphics[%natwidth=31.98cm, natheight=13.35cm, 
width=8cm, 
keepaspectratio=true]{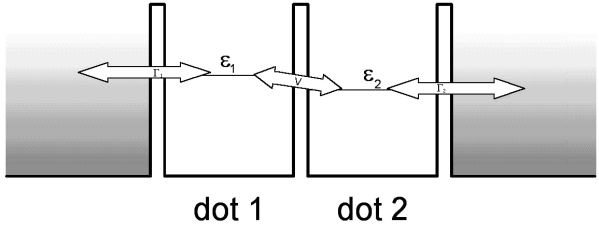}
\caption{Schematic diagram of the problem.}
\label{problem}
\end{figure}
Each dot is coupled to an environment (a triple-barrier system) as in 
Fig.~\ref{problem}.
The environment consists of a continuum of states, as would be appropriate for a
macroscopic lead. Only one state in each quantum dot is considered, which amounts 
to the assumption that the tunneling parameters connecting our two states to 
any neglected state are much less than the energy differences with that state.

The Hamiltonian of the system can be written as the sum of two terms,
%\begin{equation} \label{eq:ham}
$H=H_0 + H_c$.
%\end{equation}
The first term, $H_0$, represents the dynamics of the simple two-state system,
\begin{equation}
H_0=\varepsilon_1 d_1^\dagger d_1 + \varepsilon_2 d_2^\dagger d_2 + V(d_1^\dagger d_2 
+ d_2^\dagger d_1)\doteq \left( \begin{array}{cc} \varepsilon_1 & V \\ V & 
\varepsilon_2 \end{array} \right),
\end{equation}
where $d_i^\dagger$ creates an electron in dot $i$, and the other symbols
are defined in Fig.\ 1.
$V$ may be chosen real and positive without loss of generality.
The second term, $H_c$, reflects the 
coupling of each dot to its respective lead,
\begin{equation}
H_c=\sum_{i=1}^2
\sum_k(V_{ki} c_{ki}^\dagger d_i + V_{ki}^\ast d_i^\dagger c_{ki} + 
\varepsilon_{ki} c_{ki}^\dagger c_{ki}),
\end{equation}
where the sum on $k$ runs over the states in the continuum adjoining dot $i$, $c_{ki}^\dagger$ creates an electron in the $k$-th state of the continuum adjoining dot $i$,
$\varepsilon_{ki}$ is the energy of the $k$-th state of the appropriate 
continuum, 
and $V_{ki}$ is the tunneling parameter from the dot to the continuum.
Throughout the paper, we use units with $\hbar=1$.

In the following, 
we adopt a Green's function approach which parallels that taken
in Ref.\ \cite{sbnuc}. The retarded Green's function is defined by
\begin{equation}
G_{ij}(t)=-i\Theta(t)\langle\{d_i(t),d_j^\dagger(0)\}\rangle,
\end{equation}
where $\Theta(t)$ is the well known step function, $\{,\}$ denotes the
anticommutator, and $\langle\quad\rangle$ denotes the quantum mechanical and thermodynamic average. The retarded Green's function is the sum of the amplitude for an electron
to propagate forward in time and that for a hole to propagate backward in
time.  As such, it is a purely dynamical quantity, independent of the
occupancies of the energy levels of the system.

It is a simple matter to calculate $G_0(t)$ from $H_0$ in 
the Heisenberg picture. From this, one can find the Fourier transform 
$G_0(E)=\int_{-\infty}^{\infty} \mathrm{d} t\, G_0(t)\, e^{iEt}$, 
and finally the inverse
\begin{equation}
G_0^{-1}(E)\doteq\left( \begin{array}{cc} E-\varepsilon_1+i0^+ & -V \\ -V & 
E-\varepsilon_2+i0^+ \end{array} \right).
\end{equation}
The full Green's function can then be obtained from Dyson's equation,
\begin{equation}
G^{-1}(E)=G_0^{-1}-\Sigma.
\end{equation}
Here $\Sigma$ is the self-energy due to interactions with the continua,
\begin{equation}
\Sigma\doteq\left( \begin{array}{cc} -i\Gamma_1/2 & 0 \\ 0 & 
-i\Gamma_2/2 \end{array} \right).
\end{equation}
In the broad-band limit for the continua,
the $\Gamma_i$'s are independent of energy, and are \cite{swrabi}
\begin{equation}
\Gamma_i = 2\pi\sum_k|V_{ki}|^2 \delta(E-\varepsilon_{ki}).
\end{equation}
Application of Dyson's equation and inversion yields the total Green's function 
of the energy,
\begin{equation}
G(E) \doteq \left[\left(E-\varepsilon_1+i\frac{\Gamma_1}{2}\right) 
\left(E-\varepsilon_2+i\frac{\Gamma_2}{2}\right)-V^2\right]^{-1} %\times
\left(\begin{array}{cc} E-\varepsilon_2+i\frac{\Gamma_2}{2} & V \\ V & 
E-\varepsilon_1+i\frac{\Gamma_1}{2} \end{array} \right).
\end{equation}
$G(t)$ is then the inverse Fourier transform of $G(E)$.

Consider a single electron initially localized in dot 1 at time $t=0$. From the first row of $G(t)$, we can determine the probability $P_2$ for the electron to tunnel to dot 2 and the probability $P_1$ for the electron to remain at or return to dot 1, as functions of time,
\begin{multline}
P_1(t)=|G_{11}(t)|^2=\frac{V^2}{|\omega|^2}e^{-\overline{\Gamma}t}\Big( 
\frac{\omega_i+\Gamma'}{\Gamma'-\omega_i} e^{\omega_i t}\\
+ \frac{\Gamma'-\omega_i}{\omega_i+\Gamma'} e^{-\omega_i t} + 
\frac{i\omega_r+\Gamma'}{i\omega_r-\Gamma'} e^{i\omega_r t} + 
\frac{i\omega_r-\Gamma'}{i\omega_r+\Gamma'} e^{-i\omega_r t} \Big),
\end{multline}
\begin{equation}
P_2(t)=|G_{12}(t)|^2=\frac{2V^2}{|\omega|^2} e^{-\overline{\Gamma} t} 
(\cosh\omega_i t - \cos\omega_r t),
\end{equation}
where
\begin{equation}
\overline{\Gamma}\equiv\frac{\Gamma_1 + \Gamma_2}{2}, \qquad 
\Gamma'\equiv\frac{\Gamma_2 - \Gamma_1}{2}, 
\end{equation}
and
\begin{equation}
\omega\equiv\omega_r + i\omega_i\equiv\sqrt{4V^2 + 
(\varepsilon_2-\varepsilon_1-i\Gamma')^2}.
\label{omega}
\end{equation}
Note that the processes in which a different electron enters a dot from one of the leads, and is annihilated by $d_i(t)$, give a vanishing contribution to $G(t)$ because such processes occur with random phases and average to zero.

Some understanding of the physical meaning of the solution may be arrived at
by examining the limiting case of identical quantum dots,
i.e., $\varepsilon_1=\varepsilon_2$. In that case, we have
\begin{equation}
\omega=\sqrt{4V^2-{\Gamma'}^2},
\end{equation}
where $\omega$ must be either purely real or purely imaginary.
Working with $P_2(t)$ for algebraic simplicity, we find that if $2V>|\Gamma'|$, 
$\omega$ is purely real, and
\begin{equation}
P_2(t)=\frac{4V^2}{\omega^2}e^{-\overline{\Gamma}t}\sin^2\frac{\omega t}{2}.
\label{evo_coh}
\end{equation}
%$P_1(t)$ may be calculated at a particular time from knowledge of $P_2(t)$. 
In 
this limit, the solution exhibits the characteristic coherent behavior of Rabi 
oscillations, as one would expect from a simple two-well problem. The presence 
of the environment is visible in the exponentially decaying envelope function:
electrons leave/enter the two-state system at an overall rate $\bar{\Gamma}$.

In the case that $\varepsilon_1=\varepsilon_2$ and $2V<|\Gamma'|$,
on the other hand,
we find a purely imaginary $\omega$, and
\begin{equation}
P_2(t)=\frac{4V^2}{|\omega|^2}e^{-\overline{\Gamma}t}\sinh^2\frac{|\omega|t}{2}.
\label{evo_incoh}
\end{equation}
The hyperbolic function and the resulting lack of Rabi oscillations demonstrate 
that interdot tunneling is completely incoherent in this regime.

In the more general case that $\varepsilon_1\ne\varepsilon_2$,
 neither the circular nor 
the hyperbolic functions vanish, giving a combination of both coherent and 
incoherent behavior. For $\Gamma'=0$, however, a purely real $\omega$ is 
recovered, and the coherent evolution of
Eq.\ (\ref{evo_coh}) is obeyed, with Rabi frequency
\begin{equation}
\omega=\sqrt{4V^2+(\varepsilon_1-\varepsilon_2)^2}.
\end{equation}
This somewhat startling result is especially gratifying when viewed in the 
context of measurement theory: $\Gamma'=0$ indicates that the coupling of the 
two-well system to the environment does not distinguish between the dots, and 
thus the coherence between the dots is in no way destroyed by the macroscopic 
world.

\begin{figure}[tbp]
\includegraphics[%natwidth=14.13cm, natheight=7.78cm, 
width=6cm, keepaspectratio=true]{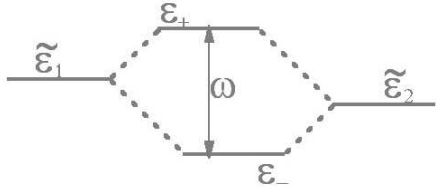}
\caption{\label{hybrid} Hybridization of the complex energy levels 
$\tilde{\varepsilon}_i$ %by the potential $V$ 
yields resonances %energy levels 
$\varepsilon_\pm$ with complex level splitting $\omega$, 
the nature of which determines the behavior of the system.}
\end{figure}

Additional insight may be gained by examining the poles of the 
Green's function, which can be interpreted as complex
energy levels of the system, or resonances \cite{ggc}. 
Defining complex energy levels
\begin{equation}
%\overset{\sim}
\tilde{\varepsilon}_i=\varepsilon_i-i\Gamma_i/2
\end{equation}
for each dot by including the local self-energy, 
the poles of $G(E)$ may be written as
\begin{equation} 
\varepsilon_\pm=\frac{\tilde{\varepsilon}_1+
\tilde{\varepsilon}_2}{2} \pm 
\sqrt{\left( \frac{\tilde{\varepsilon}_2-
\tilde{\varepsilon}_1}{2} 
\right)^2 + V^2}.
\label{eq:hyb}
\end{equation}
This is the standard formula for the hybridization of two energy levels coupled
by a matrix element $V$, except that the energy levels here are complex
(see Fig. \ref{hybrid}).
It can be understood from Dyson's equation that the poles of $G(E)$ should have this form, if the interdot tunneling is taken as the perturbation, 
rather than the coupling to the environment.  Neglecting interdot 
tunneling, the Green's function is diagonal, and consists of two
independent Breit-Wigner resonances at $\tilde{\varepsilon}_1$ and
$\tilde{\varepsilon}_2$.
The interdot tunneling operator then hybridizes these resonances
via Dyson's equation, leading to the poles displayed in Eq.\
(\ref{eq:hyb}).
%Figure~\ref{hybrid}~demonstrates this picture.
From Eqs.\ (\ref{omega}) and (\ref{eq:hyb}), it is easy to see that the 
energy difference is
\begin{equation}
\varepsilon_+ - \varepsilon_- = \omega.
\end{equation}
Thus $\omega$ can be interpreted as the complex level splitting
between the hybridized states, corresponding to a {\em complex Rabi
frequency}:  $\omega_r$ is the frequency of real Rabi oscillations,
while $\omega_i$ is the rate of incoherent tunneling.

We have presented an exactly solvable model for the tunneling dynamics
of a two-level quantum system coupled to a macroscopic environment,
appropriate to
describe a double quantum dot connected to conducting leads.
Despite its simplicity, this model displays many of the essential features
of more complicated models \cite{leg}.  
The tunneling dynamics is determined by the complex Rabi frequency given
in Eq.\ (\ref{omega}).  Our results can be interpreted in terms of 
measurement theory:  interdot coherence is suppressed to the degree
that the environment can distinguish between the two orbitals.
%We have shown exactly that the tunneling dynamics of our simple system combine 
%both coherent and incoherent components. In various limits, either the coherent
%or decoherent behavior can be eliminated. 
%The most interesting limit is when the
%coupling of each dot to the environment is identical, in which case the 
%coherence of the dots is unaffected by the environment, and the Rabi 
%oscillations are preserved.
\\
\\
\textbf{\emph{Acknowledgements}}\indent The authors
thank Ned S.\ Wingreen for useful
conversations. This work was supported in part
by NSF grants PHY0070858 and DMR0072703.


\begin{references}
\bibitem{sci95}{\sc L. Kouwenhoven}, Science {\bf 268}, 1440 (1995).
\bibitem{prl98}{\sc T. H. Oosterkamp, S. F. Godijn, M. J. Uilenreef, Y. V. 
Nazarov, N. C. van der Vaart,} and {\sc L. P. Kouwenhoven}, Phys. Rev. Lett. 
{\bf 80}, 4951 (1998).
\bibitem{nat98}{\sc T. H. Oosterkamp, T. Fujisawa, W. G. van der Wiel, K. 
Ishibashi, R. V. Hijman, S. Tarucha,} and {\sc L. P. Kouwenhoven}, Nature {\bf 
395}, 873 (1998).
\bibitem{sci98}{\sc T. Fujisawa, T. H. Oosterkamp, W. G. van der Wiel, B. W. 
Broer, R. Aguado, S. Tarucha,} and {\sc L. P. Kouwenhoven}, Science {\bf 282}, 
932 (1998).
\bibitem{sci01}{\sc M. Bayer, P. Hawrylak, K. Hinzer, S. Fafard, M. Korkusinski, 
Z. R. Wasilewski, O. Stern,} and {\sc A. Forchel}, Science {\bf 291}, 451 
(1998).
\bibitem{leg}{\sc A. J. Leggett, S. Chakravarty, A. T. Dorsey, M. P. A. Fisher, 
A. Garg,} and {\sc W. Zwerger}, Rev. Mod. Phys. {\bf 59}, 1 (1987).
\bibitem{sbnuc}{\sc C.~A.~Stafford} and {\sc B.~R.~Barrett}, Phys.
Rev. C {\bf 60}, 053105 (1999).
\bibitem{swrabi}{\sc C.~A.~Stafford} and {\sc N.~S.~Wingreen}, Phys. Rev. Lett. 
{\bf 76}, 1916 (1996).
\bibitem{ggc} {\sc G. Garc\'{\i}a-Calder\'on, R. Romo,} and {\sc A. Rubio}, Phys. Rev. B
{\bf 47}, 9572 (1993).
\end{references}
\end{document}